\documentclass[runningheads]{llncs}
\usepackage[T1]{fontenc}
\usepackage{hyperref}

\usepackage{graphicx}
\usepackage{amsmath,amssymb, mathtools}
\usepackage{bm}
\usepackage[misc,geometry]{ifsym}
\usepackage{hyperref}
\begin{document}
%
\title{Reducing Positional Variance in Cross-sectional Abdominal CT Slices with Deep Conditional Generative Models}
\titlerunning{Reducing Positional Variance in Cross-sectional Abdominal CT Slices}

\author{Xin Yu\inst{1*}(\Letter) \and
Qi Yang\inst{1*} \and
Yucheng Tang\inst{2} \and
Riqiang Gao\inst{1} \and
Shunxing Bao\inst{1} \and
Leon Y. Cai \inst{3} \and
Ho Hin Lee\inst{1} \and
Yuankai Huo\inst{1} \and
Ann Zenobia Moore \inst{4}\and
Luigi Ferrucci \inst{4} \and
Bennett A. Landman\inst{1,2,3}}
\authorrunning{X. Yu et al.}
\institute{Computer Science, Vanderbilt University, Nashville, TN, USA \\
\email{xin.yu@vanderbilt.edu}\\
\and
Electrical and Computer Engineering,  Vanderbilt University, Nashville, TN, USA \and
Biomedical Engineering, Vanderbilt University, Nashville, TN, USA \and
National Institute on Aging, Baltimore, MD, USA}
\maketitle              
\begin{abstract}
2D low-dose single-slice abdominal computed tomography (CT) slice enables direct measurements of body composition, which are critical to quantitatively characterizing health relationships on aging. However, longitudinal analysis of body composition changes using 2D abdominal slices is challenging due to positional variance between longitudinal slices acquired in different years. To reduce the positional variance, we extend the conditional generative models to our C-SliceGen that takes an arbitrary axial slice in the abdominal region as the condition and generates a defined vertebral level slice by estimating the structural changes in the latent space. Experiments on 1170 subjects from an in-house dataset and 50 subjects from BTCV MICCAI Challenge 2015 show that our model can generate high quality images in terms of realism and similarity. External experiments on 20 subjects from the Baltimore Longitudinal Study of Aging (BLSA) dataset that contains longitudinal single abdominal slices validate that our method can harmonize the slice positional variance in terms of muscle and visceral fat area. Our approach provides a promising direction of mapping slices from different vertebral levels to a target slice to reduce positional variance for single slice longitudinal analysis. The source code is available at: \url{https://github.com/MASILab/C-SliceGen}.
\keywords{Abdominal Slice Generation \and Body Composition \and Longitudinal Data Harmonization.}
\end{abstract}

\def\thefootnote{*}\footnotetext{Equal contribution}

\section{Introduction}
\label{Introduction}
 \begin{figure}
\includegraphics[width=\textwidth]{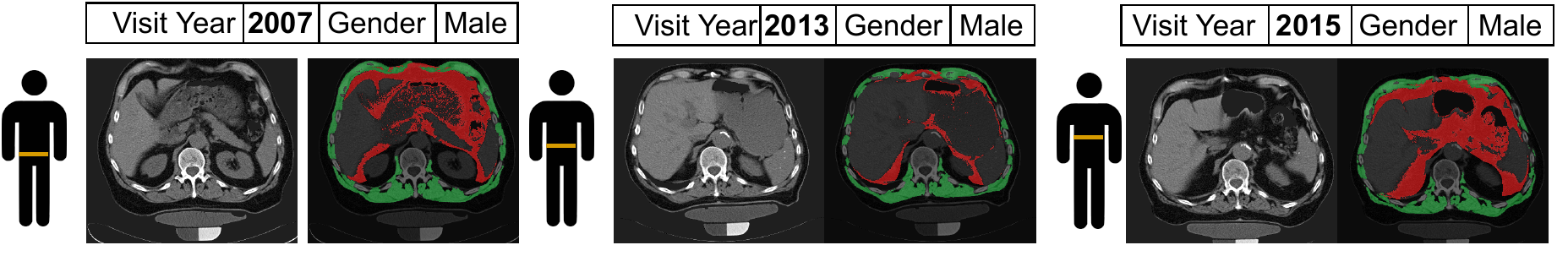}
\caption{Longitudinal slices acquired at different axial cross-sectional positions. The figure shows a single abdominal slice being acquired for the same subject in different years. The orange line represent the given CT axial position. The red and green masks represent the visceral fat and muscle, respectively. The area of two masks varies largely due to the slice positional variance.}
\label{fig:fig1}
\end{figure}
 Body composition describes the percentage of fat, muscle and bone in the human body \cite{kuriyan2018body} and can be used to characterize different aspects of health and disease, including sarcopenia \cite{ribeiro2014sarcopenia}, heart disease\cite{de2011body} and diabetes \cite{solanki2015body}. One widely used method to assess body composition is computed tomography body composition (CTBC) \cite{andreoli2016body}. 2D low-dose axial abdominal single-slice computed tomography (CT) is preferred over 3D CT to reduce unnecessary radiation exposure \cite{kuriyan2018body}. However, difficulty in positioning cross-sectional locations leads to challenges acquiring the 2D slices at the same axial position (vertebral level). For instance, in the clinical setting, patients who visit hospitals in the different years can have varied vertebral level abdominal slices being scanned (Fig.~\ref{fig:fig1}). This causes significant variation in the organs and tissues captured. Since the specific organs and tissues scanned are highly associated with measures of body composition, increased positional variance in 2D abdominal slices makes body composition analyses difficult. To the best of our knowledge, no method has been proposed to tackle the 2D slice positional variance problem.




We aim to reduce the positional variance by synthesizing slices to a target vertebral level. In other contexts, image registration would be used to correct pose/positioning. However, such an approach cannot address out of plane motion with 2D acquisitions. 
Recently, deep learning based generative models have shown superior results in generating high-quality and realistic images. The generative model can solve the registration limitation by learning the joint distribution between the given and target slices. Variational autoencoders (VAEs), a class of generative models, encode inputs to an interpretable latent distribution capable of generating new data \cite{kingma2013auto}. Generative adversarial networks (GANs) contain two sub-models: a generator model that aims to generate new data and a discriminator to distinguish between real and generated images.
VAEGAN \cite{larsen2016autoencoding} incorporates GAN into a VAE framework to create better synthesized images. In the original VAEs and GANs, the generated images cannot be manipulated. Conditional GAN (cGAN) \cite{mirza2014conditional} and conditional VAE (cVAE) \cite{sohn2015learning} tackle this by giving a condition to generate specific data. However, the majority of these conditional methods require the target class, semantic map, or heatmap \cite{de2019conditional} in the testing phase, which is not applicable in our scenario where no direct target information is available.

We posit that by synthesizing an image at a pre-defined vertebral level with an arbitrary abdominal slice, generated slices will consistently localize to the target vertebral level and the subject-specific information derived from the conditional image such as body habitus will be preserved. Inspired by \cite{henderson2021unsupervised,de2019conditional,tang2021pancreas}, we propose the Conditional SliceGen (C-SliceGen) based on VAEGAN. C-SliceGen can generate subject-specific target vertebral level slice given an arbitrary abdominal slice as input. To ensure the correctness of our model, we train and validate our method first on an inhouse 3D volumertric CT dataset and the BTCV MICCAI Challenge 2015 3D CT dataset \cite{landman2015miccai} where target slices are acquired and used as ground truth to compare with generated slices. SSIM, PSNR, and LPIPS are used as evaluation metrics. Our experiments show that our model can capture positional variance in the generated realistic image. Moreover, we apply our method to the Baltimore Longitudinal Study of Aging (BLSA) single-slice dataset \cite{ferrucci2008baltimore}. By computing body composition metrics on synthesized slices, we are able to harmonize the longitudinal muscle and visceral fat area fluctuations brought by the slices positional variation. 

Our contributions are three-fold: (1) we propose C-SliceGen to successfully capture positional variance in the same subject; (2) the designed generative approach implicitly embeds the target slice without requiring it during the testing phase; and (3) we demonstrate that the proposed method can consistently harmonize the body composition metrics for longitudinal analysis.

\section{Method}
\subsection{Technical Background}
\label{problem}

\noindent {\bf VAE} VAEs can be written in probabilistic form as $P(x)=P(z)P(x|z)$, where x denotes the input images and z denotes the latent variables. The models aim to maximize the likelihood $p(x)=\int p(z)p_\theta(x|z)dz$, where $z\sim N(0,1)$ is the prior distribution, $p_\theta(x|z)dz$ is the posterior distribution and $\theta$ is the decoder parameters. However, it is intractable to find decoder parameters $\theta$ to maximize the log likelihood. Instead, VAEs optimize encoder parameters $\phi$, by computing $q_\phi(z|x)$ to estimate $p_\theta(x|z)$ with the assumption that $q_\phi(z|x)$ is a Gaussian distribution whose $\mu$ and $\sigma$ are the outputs of the encoder. VAEs train the encoder and decoder jointly to optimize the Evidence Lower Bound (ELBO), 
\begin{equation}
L_{VAE}(\theta,\phi,x,y) = E[\log p_{\theta}(x|z)] - D_{KL}[q_{\phi}(z|x)||p_{\theta}(z)],
\end{equation}
where $E[\log p_{\theta}(x|z)]$ represent the reconstruction loss and the KL-divergence encourages the posterior estimate to approximate the prior $p(z)$ distribution. Samples can be generated from the normal distribution $z \sim N(0,1)$ and fed into the decoder to generate new data during testing time. cVAEs add flexibility to the VAEs and can also be trained by optimizing the ELBO. 

\subsubsection{WGAN-GP} Wasserstein GAN (wGAN) with gradient penalty\cite{gulrajani2017improved} is an extension of GAN that improves stability when training the model whose loss function can be written as:
\begin{figure}
\includegraphics[width=\textwidth]{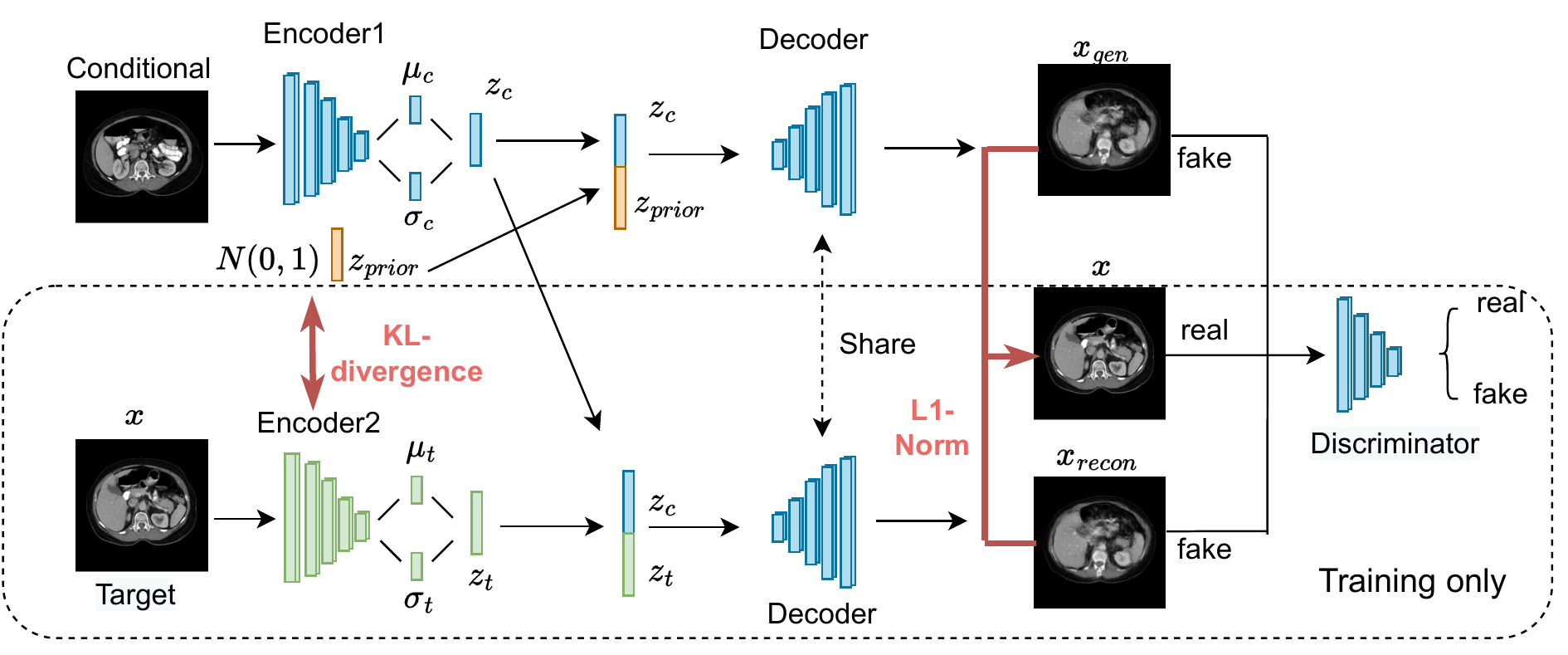}
\caption{Overall pipeline. Conditional images, arbitrary slices in the abdominal region, are the input images for both training and test phase. Target images ($x$) are the ground truth for the reconstruction and generation process that only exist in the training phase. $z_c$, $z_t$ and $z_{prior}$ are the latent variables derived from conditional images, target images, and the normal Gaussian distribution, respectively. $x_{gen}$ and $x_{recon}$ serve as fake images and target images serve as real images for the discriminator.}
\label{fig:fig2}
\end{figure}
\begin{equation}
L_{GAN} = \mathbb{E}_{\tilde{x}\sim\mathbb{P}_g}[D(\tilde{x})] +\mathbb{E}_{x\sim \mathbb{P}_r}[D(x)] + \lambda \mathbb{E}_{\hat{x}\sim \mathbb{P}_{\hat{x}}}[(||\bigtriangledown_{\hat{x}}D(\hat{x})||_2 - 1 )^2] ,
\label{equ:equ2}
\end{equation}
where $\mathbb{P}_g$ is the model distribution,  $\mathbb{P}_r$ is the data distribution and $\mathbb{P}_{\hat{x}}$ is the random sample distribution.

Although the current VAE/GAN based methods have been successful in generating samples, no existing method found can be directly applied to our task.
\subsection{C-SliceGen}

In our scenario, acquired slice can be in any vertebral level within the abdominal region. The goal is to use these arbitrary slices to synthesize a new slice at a pre-defined target vertebral level. The overall method is shown in Fig.~\ref{fig:fig2}. There are two encoders, one decoder and one discriminator. The arbitrary slice is the conditional image for the model, which provides subject-specific information such as organs shape and tissue localization. We assume this information remains interpretable after encoding to latent variables $z_c$ by encoder1.

 \noindent {\bf Training}
During training phase, we have the sole target slice for each subject. We select the most similar target slice in terms of organ/tissue structure and appearance for all subjects. The target slice selection method will be covered in the following section. We assume all target slices $x$ to have similar organ/tissue structure and appearance which are preserved in the latent variables $z_t$, whose distribution can be written as $q_\phi(z_t|x)$. By concatenating the latent variables $z_c$ and $z_t$, we combine the target slice organ/tissue structure and appearance with given subject-specific information. This combined latent variable encourages the decoder to reconstruct the target slice for the given subject. We regularize this reconstruction step by computing the L1-Norm between the target slice ($x$) and the reconstructed slice ($x_{recon}$), denoted as $L_{recon} = \lVert\bm{x} - \bm{x}_{recon}\rVert$. In testing phase, however, no target slices are available. To solve this problem, we follow the similar practice in VAEs: We assume $q_\phi(z_t|x)$ is a Gaussian distribution parameterized by the outputs of encoder2 $\mu_t$ and $\sigma_t$, and encourage $q_\phi(z_t|x)$ to be close to the $z_{prior} \sim N(0,1)$ by optimizing the KL-divergence, written as: 
\begin{equation}
L_{KL}=\frac{1}{2}\sum_{k=1}^K(1+\log(\sigma_k^2) - \mu_k^2 - \sigma_k^2),
\end{equation}
where $K$ is latent space dimension. To further constrain this KL term and to mimic the test phase image generation process, we concatenated $z_c$ with $z_{prior}$ as another input of the decoder for target slices generation. These generated slices are denoted as $x_{gen}$. We compute the L1-Norm between the generated image ($x_{gen}$) and target image ($x$), denoted as $L_{gen}=\lVert\bm{x} - \bm{x}_{gen}\rVert$. The total loss function of the aforementioned steps is written as:
\begin{equation}
L_{cVAE}=L_{recon} + L_{gen} + L_{KL},
\label{equ:equ4}
\end{equation}

However, maximize likelihood function is inherently a difficult problem which can cause blurry generated images. GANs on the other hand increase the image quality in an adversarial manner. Following \cite{larsen2016autoencoding}, we combine GAN with our VAE model. The generated image and reconstructed image both serve as fake images, and the target images serve as the real images for the discriminator to perform classification. The decoder serves as the generator. By sharing the same parameters for the generated and reconstructed images, the GAN loss adds another constraint to force them to be similar. The total loss function of our proposed C-SliceGen can be written as:
\begin{equation}
L = L_{cVAE} + \beta L_{GAN},
\label{equ:equ5}
\end{equation}
where $\beta$ is a weighting factor that determines the adversarial regularization. 

\noindent {\bf Testing} 
During testing, given a conditional image as input, the encoded latent variable $z_c$ is concatenated with $z_{prior}$ sampled from a normal Gaussian distribution and is fed into the decoder to generate the target slice.

\subsection{Target Slice Selection}
\label{problem}
It is not straightforward to select similar target slices for each subject since body composition and organ structure are subject-sensitive. We adopt two approaches: (1) Select the slices that have the most similar body part regression (BPR) \cite{tang2021body} score as the target slices across subject since BPR is efficacious in locating slices. (2) Select a slice from a reference subject as the reference target slice. Registering axial slice of each subject's volume to the reference target slice and identifying the slice with the largest mutual information \cite{cover1999elements} as the subject target slice. 

\section{Experiments and Results}
\subsection{Dataset}
\label{dataset}
The models are trained and validated on a large dataset containing 1170 3D Portal Venous CT volumes from 1170 de-identified subjects under Institutional Review Board (IRB) protocols. Each CT scan is quality checked for normal abdominal anatomy. The evaluations are performed on the MICCAI 2015 Multi-Atlas Abdomen Labeling Challenge dataset which contains 30 and 20 abdominal Portal Venous CT volumes for training and testing, respectively. We further evaluate our method's efficacy on reducing positional variance for longitudinal body composition analysis on 20 subjects from the BLSA non-contrast single slice CT dataset. Each subject has either 2 or 3 visits for the past 15 years.

\subsection{Implementation Details and Results}
\noindent {\bf Metrics}
We quantitatively evaluate our generative models C-SliceGen with different $\beta$ (Eq.~\ref{equ:equ5}) and the two different target slice selection approaches using three image quality assessment metrics: Structural Similarity
Index (SSIM) \cite{wang2004image}, Peak Signal-to-Noise Ratio (PSNR)\cite{hore2010image}, and Learned Perceptual Image Patch Similarity (LPIPS) \cite{zhang2018perceptual}. 


 \noindent {\bf Training \& Testing}
All 3D volumes undergo BPR to ensure a similar Field of View (FOV). The 2D axial CT scans have image sizes of $512 \times 512$ and are resized to $256 \times{256}$ before feeding into the models.The data are processed with soft-tissue CT window range [-125, 275] HU and rescaled to [0.0,1.0] to facilitate training. The proposed methods are implemented using Pytorch. We use Adam optimizer with a learning rate of $1e-4$ and weight decay of $1e-4$ to optimize the total loss of the network. We adapt the encoder, decoder and discriminator structures in \cite{gao2021lung,li2020learning} to fit our input size. Shift, rotation and flip are used for the online data augmentation.\begin{table}[]
\centering
\caption{Quantitative results on the in-house test set and BTCV test set. Registration: target slice selected using registration, BPR: target slice selected using BPR score alone. $\beta$ represent the $\beta$ in Eq.~\ref{equ:equ5}.}
\label{tab:my-table}
\begin{tabular}{p{3.5cm}|p{2.5cm}p{2.5cm}p{2.5cm}}
\hline
Method  & SSIM $\uparrow$ & PSNR  $\uparrow$ & LPIPS $\downarrow$ \\ \hline
\multicolumn{4}{c}{the in-house dataset} \\ \hline
$\beta = 0$, Registration & 0.636 & 17.634 & 0.361 \\
$\beta = 0$, BPR & 0.618 & 16.470 & 0.381 \\
$\beta = 0.01$, Registration & 0.615 & 17.256  & 0.209 \\
$\beta = 0.01$, BPR & 0.600 & 16.117 & 0.226 \\ \hline
 \multicolumn{4}{c}{the BTCV dataset} \\ \hline
$\beta = 0$, Registration & 0.603 & 17.367 & 0.362 \\
$\beta = 0$, BPR & 0.605 & 17.546 & 0.376 \\
$\beta = 0.01$, Registration & 0.583 & 16.778 & 0.208 \\
$\beta = 0.01$, BPR & 0.588 & 16.932 & 0.211 \\ \hline

\end{tabular}
\end{table} 2614 slices from in-house 83 subjects are used for testing. The results are shown in Table.~\ref{tab:my-table}.

\begin{figure}
\centering
\includegraphics[width=\textwidth]{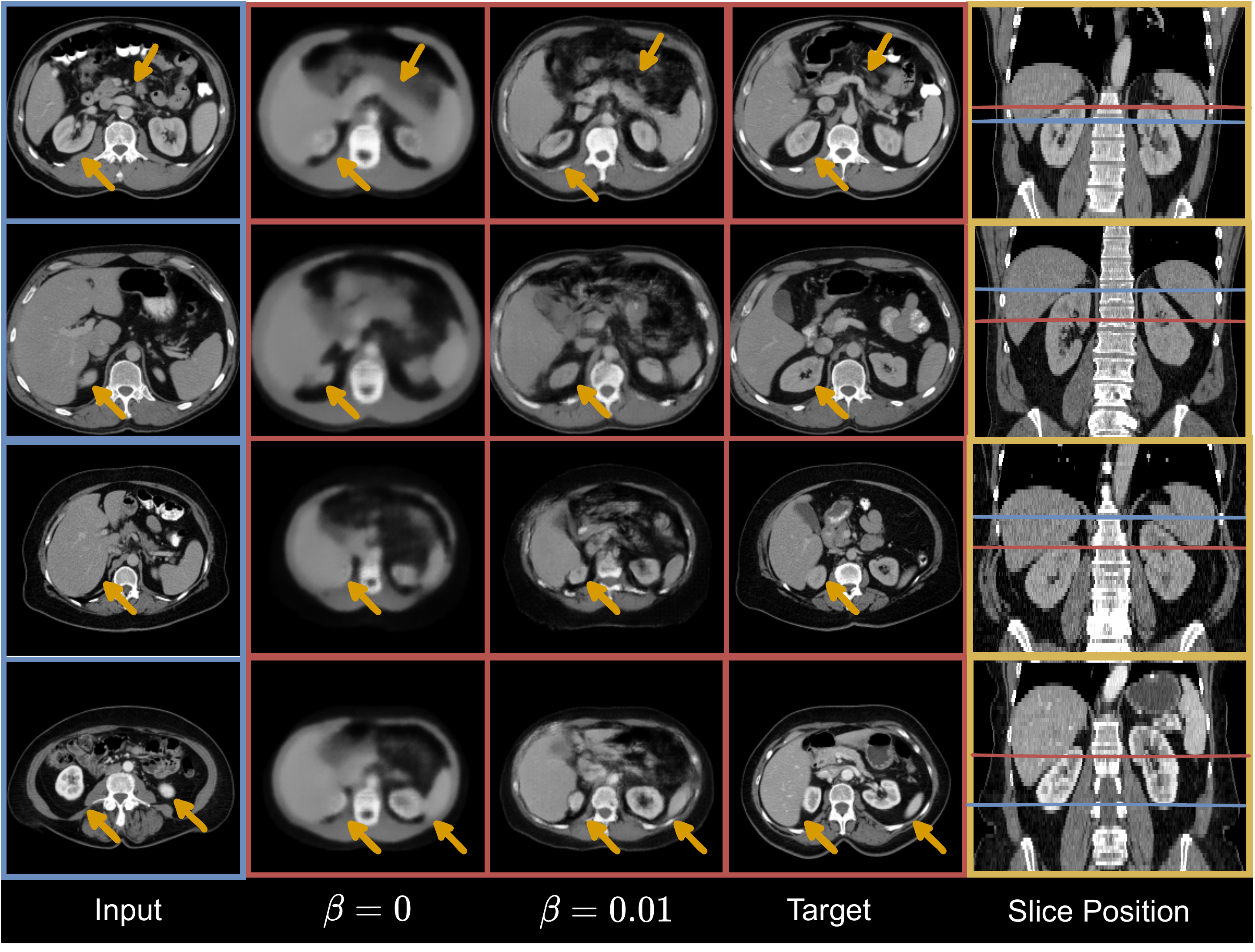}
\caption{Qualitative results on the BTCV test set. The image with blue bounding box represent the input slice while the image with red bounding box represents the model results and target slice. In the rightmost column, axial location of the input and target slices are marked with blue line and red line, respectively.}
\label{fig3}
\end{figure}

\noindent {\bf BTCV Evaluation}
Before evaluating on the BTCV test set, the models are fine-tuned on the BTCV training set to minimize the dataset domain gap with the same training settings except that the learning rate is reduced to $1e-5$. The split of train/validation/test is 22/8/20. The quantitative results and qualitative results are shown in Table.~\ref{tab:my-table} and Fig.~\ref{fig3}, respectively.

\noindent {\bf BLSA Evaluation}
In the BLSA 2D abdominal dataset, each subject only take one axial abdominal CT scan each year. Therefore, there is no ground truth (GT) for target slice generation. Instead of directly comparing the generated images with the GT, we evaluate the model performance on reducing variance in body compositional areas brought by the cross-sectional variance of CT scans in longitudinal data, specifically muscle and visceral fat. We feed the BLSA data into our C-SliceGen model and resize the generated images to the original size of $512 \times 512$. Both the real and generated images are fed into a pre-trained UNet \cite{ronneberger2015u} for muscle segmentation and a pre-trained Deeplab-v3 \cite{florian2017rethinking} to identify visceral fat by inner/outer abdominal wall segmentation.  For the inner/outer wall segmentation, we find that the model usually fails to exclude the retroperitoneum in the inner abdominal wall for both real and fake images. Retroperitoneum is an anatomical area located behind the abdominal cavity including the left/right kidneys and aorta, often with poorly visualized boundaries.We conduct human assessment on all the results from both the real and generated images to ensure the retroperitoneum is correctly segmented. The adipose tissue of each image is segmented using fuzzy c-means
\cite{bezdek1984fcm,tang2020prediction} in an unsupervised manner. We segment the visceral fat by masking the adipose tissue with the inner abdominal wall. Fig.~\ref{fig4} shows the spaghetti plots of the muscle and visceral fat area changes among 2 or 3 visits from 20 subjects before and 
after harmonization.

\section{Discussion and Conclusion}

According to the qualitative results shown in Fig.~\ref{fig3}, our generated images with $\beta = 0.01$ are realistic and similar to the target slices. Specifically, the results show that our model can generate target slices regardless of whether the conditional slice is at an upper, lower, or similar vertebral level.  Comparing the results between $\beta = 0$ and $\beta =0.01$, the images indicate that the adversarial regularization helps improve image quality significantly.\begin{figure}
\centering
\includegraphics[width=\textwidth]{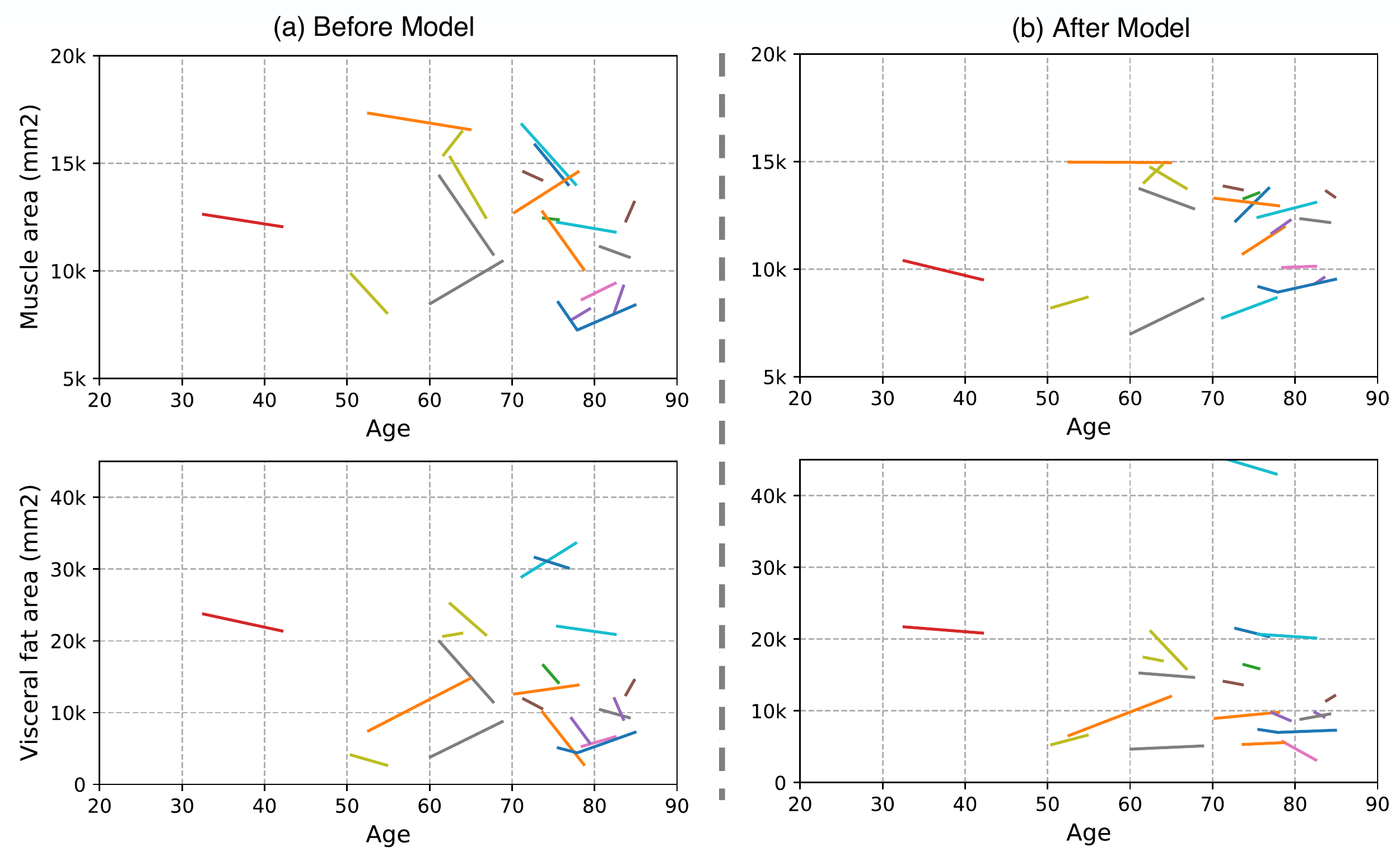}
\caption{Spaghetti plot of muscle and visceral fat area longitudinal analysis with 20 subjects from the BLSA dataset. (a) muscle and visceral fat area derived from the original abdominal slices. (b) the corresponding metrics derived from synthesized slices with our model. Each line corresponds to the measurements across visits for one subject. After harmonization, variance in these measures for the population is decreased.}
\label{fig4}
\end{figure}  This is consistent with the LPIPS results in Table.~\ref{tab:my-table}.However, the human qualitative assessment and LPIPS differ from the SSIM and PSNR as shown in Table.~\ref{tab:my-table} where SSIM and PSNR have higher scores with $\beta = 0$ on both dataset. This supports that SSIM and PSNR score may not fully represent a human perceptional assessment \cite{ledig2017photo,almalioglu2020endol2h}. As for the longitudinal data harmonization, according to Fig.~\ref{fig4}, before applying our model, both muscle and visceral fat area have large fluctuations. These fluctuations have been reduced after mapping the slices to a similar vertebral level with our model C-SliceGen.

As the first work to use one abdominal slice to generate another slice, our approach currently has several limitations. (1) In most cases, the model is able to identify the position of each organ, but shape and boundary information are not well preserved. (2) It is hard to synthesize heterogeneous soft tissues such as the colon and stomach. (3) There is domain shift when the model trained on Portal Venous phase CT is applied to CT acquired in other phases 
such as the non-contrast BLSA data.

In this paper, we introduce our C-SliceGen model that conditions on an arbitrary 2D axial abdominal CT slice and generates a subject-specific slice at a target vertebral level. Our model is able to capture organ changes between different vertebral levels and generate realistic and structurally similar images. We further validate our model's performance on harmonizing the body composition measurements fluctuations introduced by positional variance on an external dataset. Our method provides a promising direction for handling imperfect single slice CT abdominal data for longitudinal analysis.

\noindent {\bf Acknowledgements}
This research is supported by NSF CAREER 1452485, 2040462 and the National Institutes of Health (NIH) under award numbers R01EB017230, R01EB006136, R01NS09529, T32EB001628, 5UL1TR002243-04, 1R01MH121620-01, and T32GM007347; by ViSE/VICTR VR3029; and by the National Center for Research Resources, Grant UL1RR024975-01, and is now at the National Center for Advancing Translational Sciences, Grant 2UL1TR000445-06. This research was conducted with the support from the Intramural Research Program of the National Institute on Aging of the NIH. The content is solely the responsibility of the authors and does not necessarily represent the official views of the NIH. The identified datasets used for the analysis described were obtained from the Research Derivative (RD), database of clinical and related data. The inhouse imaging dataset(s) used for the analysis described were obtained from ImageVU, a research repository of medical imaging data and image-related metadata. ImageVU and RD are supported by the VICTR CTSA award (ULTR000445 from NCATS/NIH) and Vanderbilt University Medical Center institutional funding. ImageVU pilot work was also funded by PCORI (contract CDRN-1306-04869). 

%
%
%
\bibliographystyle{splncs04}
%
\bibliography{reference}

\end{document}